\documentclass[11pt]{article}
\usepackage{geometry}
\usepackage{graphicx}
\usepackage{amssymb}
\usepackage{epstopdf}
\usepackage{tabularx}
\usepackage{placeins}
\usepackage{array}

\newcolumntype{M}[1]{>{\centering\arraybackslash}m{#1}}
\usepackage[hidelinks]{hyperref}

\title{Demographics and discussion\\influence views on algorithmic fairness}
\author{Emma Pierson\footnote{Thanks to Ashi Agrawal, Jackie Basu, Sam Corbett-Davies, Shengwu Li, Dana{\"e} Metaxa-Kakavouli, Bryan McCann, Chris Olah, Leah Pierson, and Nat Roth for thoughtful comments and assistance, and to all survey participants. Feedback is welcome at emmap1@cs.stanford.edu. Replication data, slides, and code are available at \url{https://github.com/epierson9/algorithmic_fairness_survey}.}}
\date{} 

\begin{document}
\maketitle

\abstract{The field of algorithmic fairness has highlighted ethical questions which may not have purely technical answers. For example, different algorithmic fairness constraints are often impossible to satisfy simultaneously, and choosing between them requires value judgments about which people may disagree. Achieving consensus on algorithmic fairness will be difficult unless we understand why people disagree in the first place. Here we use a series of surveys to investigate how two factors affect disagreement: \emph{demographics} and \emph{discussion}.  First, we study whether disagreement on algorithmic fairness questions is caused partially by differences in \emph{demographic backgrounds}. This is a question of interest because computer science is demographically non-representative. If beliefs about algorithmic fairness correlate with demographics, and algorithm designers are demographically non-representative, decisions made about algorithmic fairness may not reflect the will of the population as a whole. We show, using surveys of three separate populations, that there are gender differences in beliefs about algorithmic fairness. For example, women are less likely to favor including gender as a feature in an algorithm which recommends courses to students if doing so would make female students less likely to be recommended science courses. Second, we investigate whether people's views on algorithmic fairness can be changed by \emph{discussion} and show, using longitudinal surveys of students in two computer science classes, that they can.}

\section{Introduction}

The rapidly growing field of algorithmic fairness has produced many ethical dilemmas and little consensus about how to resolve them. For example, it has been repeatedly shown \cite{chouldechova2017fair, corbett2017algorithmic, kleinberg2016inherent,pleiss2017fairness} that different conceptions of fairness cannot all be achieved simultaneously, both in theory and in real data. Illuminating these conflicts is invaluable, but thus far there is little agreement about what to do about them. 

Disagreement likely arises in part because the core debates in algorithmic fairness are philosophical, not just technical. Building a classifier that can predict criminal recidivism with high accuracy is a relatively straightforward technical problem with a clearly defined success metric. But determining how to weigh maximizing accuracy against minimizing disparities requires value judgments about which people may disagree. Studies of opinions about algorithmic fairness indicate substantial disagreement \cite{grgic2018human,grgic2016case} about which features are fair to use. A recent review of the ethical considerations in data science argues, with respect to algorithmic fairness, that ``Favoring certain fairness properties over others could just as well have reflected a difference in values''  \cite{barocas2017engaging}. 

While much effort has been devoted to proposing and arguing in favor of new fairness properties, it will be hard to achieve consensus unless we understand why people disagree about them in the first place. That question has received relatively little quantitative study, and is the question we consider here. A concurrent study, \cite{grgic2018human}, argues that disagreement over whether a feature is fair to use in criminal risk prediction can be attributed to differences in how people perceive ``latent properties'' of the feature. Here we study how two factors affect disagreement on algorithmic fairness: \emph{demographics} and \emph{discussion}. 

Previous work has shown that people's moral judgments correlate with their demographic traits \cite{womenmenmoralityandethics, fumagalli2010gender}, as do beliefs about the ethics of self-driving cars \cite{bonnefon2016social} and beliefs about fairness and discrimination \cite{loughlin2000barriers, parker2016views}. These findings raise the question of whether judgments about algorithmic fairness also correlate with demographic traits. This is a question of particular interest because computer science is extremely demographically skewed \cite{landivar2013disparities}. If beliefs about algorithmic fairness correlate with demographics, and computer scientists are demographically skewed, decisions made about algorithmic fairness may not reflect what the population as a whole would want. To our knowledge, however, statistically significant demographic differences in beliefs about algorithmic fairness have not been previously shown. 

Here we show that demographics do predict beliefs about algorithmic fairness. Using surveys of three separate populations (one recruited using social media; one recruited using Google Consumer Surveys; and one of students in two undergraduate computer science classes) we find that gender predicts people's beliefs on several questions of algorithmic fairness. For example, women are less likely to favor including gender as a feature in an algorithm which recommends courses to students if doing so would make female students less likely to be recommended science courses. 

We then investigate how disagreement on algorithmic fairness is affected by discussion, by surveying students in two undergraduate computer science classes both before and after an hour-long lecture and discussion on algorithmic fairness. We find that more than 90\% of students report their views changed at least slightly. For example, students became more likely to support the use of algorithms in the criminal justice system and more likely to favor algorithmic transparency. However, heterogeneity in student views does not consistently decrease after discussion. Altogether, our results suggest that consensus on issues of algorithmic fairness may be difficult to achieve. Demographic differences contribute to disagreement, and though discussion can change views, it does not necessarily increase consensus. 

\section{Methods}

We designed a survey with four algorithmic fairness opinion questions; the full text of survey questions is available at \url{https://github.com/epierson9/algorithmic_fairness_survey/}. In brief, we presented respondents with the following situations: 

\begin{enumerate}
\item A computer program is used by an education company to recommend courses to students. Should the program use gender to recommend courses to students if this a) increases the accuracy of recommendations, making it more likely students will sign up for courses, but b) makes it less likely that women will be recommended science classes? 
\item A computer program predicts whether criminal defendants will commit another crime. This program does not use race to make decisions, but it is much more likely to rate a black defendant as high risk than a white defendant, and this is true even if neither defendant will go on to commit another crime. However, removing these disparities requires reducing the accuracy of the computer program, which could increase crime rates or increase the number of low-risk defendants who are unnecessarily jailed. Should the program be made as accurate as possible even if this creates large racial disparities, or should we make sure the algorithm produces no racial disparities, even if that produces a less accurate computer program? 
\item There is evidence that computer programs can decide more accurately than human judges how likely a defendant is to commit another crime. But some have argued that such decisions ought to be left to human judges, since computers can never fully take individual factors into account. Should we use computer programs rather than human judges to make these decisions?
\item Should private companies who design algorithms used in the criminal justice system be allowed to keep the details of their algorithms secret? 
\end{enumerate}

Respondents answered all questions on a scale from 1-7. We also asked respondents for their gender, academic background (computer science, other science or math, or humanities), and race/ethnicity. We surveyed three populations.  

\begin{enumerate}
\item \textbf{Social media survey:} We distributed the survey to an initial discovery sample through social media channels and email lists. In total 163 respondents filled out our survey completely and identified as male or female: 91 males and 72 females; 49 from computer science, 37 from other science/math, and 77 from humanities. 

\item \textbf{Google Consumer Surveys:} Because surveys conducted social media study a selected sample and our initial sample size was somewhat small, we surveyed a larger, more representative sample using Google Consumer Surveys \cite{mcdonald2012comparing} (GCS). Due to space and cost limitations on the GCS platform, we could disseminate only the ``gender in course recommendations'' opinion question. (We chose this question because a) it showed a replicable effect size and b) it was easy to explain in a small amount of space, increasing the probability that respondents would understand it.) In total our data included 573 respondents identified as male or female: 291 males and 282 females. 

\item \textbf{Computer science undergraduates:} To identify a population particularly relevant to decision-making in algorithmic fairness, we surveyed students in two undergraduate computer science classes that focused on ethical and social issues in computer science\footnote{One class was much larger than the other, so our results are primarily driven by that class; nonetheless, for completeness, we include data from both classes in our analysis.}. In both classes, the surveys were conducted as part of a lecture on algorithmic fairness with permission from instructors. The lecture was structured as follows: 
\begin{itemize}
\item To provide context for discussion, students read ProPublica's ``Machine Bias: There's software used across the country to predict future criminals. And it's biased against blacks'' prior to class \cite{angwin2016machine}. This article, which was both widely read and controversial, discussed the use of a criminal risk prediction algorithm, COMPAS, which produced much higher false positive rates for black defendants than white defendants, and much lower false negative rates for white defendants than black defendants. Subsequent analyses demonstrated that while COMPAS failed to satisfy ProPublica's definition of fairness, it satisfied other definitions of fairness, and that the definitions in general could not be simultaneously satisfied \cite{chouldechova2017fair, corbett2016computer, kleinberg2016inherent}.
\item Students completed the survey described above, with an additional question asking whether they thought the COMPAS algorithm discussed in the ProPublica article was unfair.
\item The author gave an hour-long lecture discussing issues in algorithmic fairness with opportunities for questions and discussion. (Lecture slides are available at: \url{https://github.com/epierson9/algorithmic_fairness_survey/}.) 
\item Students completed the same survey as before class, with one additional question asking whether their views had changed and if so how.
\end{itemize}

Each student generated a unique ID to allow tracking of responses between the pre- and post-class surveys. After filtering for complete responses, we were left with 22 female respondents and 65 male respondents; 61 from computer science, 23 from other science/math, and 3 from humanities. 

\end{enumerate}

Several caveats ought to be kept in mind in interpreting our results. First, our samples were not sufficiently racially representative to allow us to assess racial discrepancies in responses; similarly, we did not have a large enough sample size to assess non-binary gender identities. A second caveat is that, although we had to present the questions as one-dimensional scales to allow for comparable numerical responses, the questions had nuances which could not always be captured along a single dimension, a fact some respondents commented on. 

\section{Results}

\subsection{Do demographic differences contribute to disagreement?}

\FloatBarrier

\begin{table}[h!]
\centering
\scriptsize
\begin{tabular}{|p{5.5cm}|M{1.8cm}|M{1.3cm}|M{1.3cm}|M{1.3cm}|} \hline 
{} & Social Media &         GCS &  Pre-class & Post-class \\ \hline 
Use gender in course recommendations even if it reduces women in science classes? &  2.9, 1.7*** $p=3.3 * 10^{-5}$ &  3.0, 2.5** $p=.004$ &  2.1, 1.4* $p=.02$ &  2.5, 1.6* $p=.02$ \\ \hline
Reduce racial disparities in criminal risk prediction at expense of accuracy?     &  3.9, 5.0*** $p=4.2 * 10^{-4}$ &                    - &  4.4, 5.3* $p=.04$ &   4.4, 5.0 $p=.11$ \\ \hline
Use computer algorithms (as opposed to human judges) in criminal justice at all?  &     4.2, 3.7 $p=.06$ &                    - &   3.6, 2.8 $p=.05$ &   4.6, 3.9 $p=.07$ \\ \hline
Allow companies to keep details of criminal justice algorithms secret?            &     2.4, 2.4 $p=1.0$ &                    - &   2.2, 1.9 $p=.33$ &   1.6, 1.3 $p=.11$ \\ \hline
Is the COMPAS algorithm unfair?                                                   &                     - &                    - &   5.3, 5.5 $p=.48$ &   5.2, 4.9 $p=.27$ \\ \hline
\end{tabular}
\caption{Gender differences for all questions and survey populations. Each cell reports the mean answer for male respondents followed by the mean answer for female respondents (on a scale of 1-7). Higher numbers indicate more affirmative answers to the question. Stars denote statistical significance (two-tailed t-test, $*:p < .05, **: p < .01$, *** : $p < .001$)} 
\label{tab:combined_gender_results}
\end{table}

\begin{figure}[h!]
\centering
\includegraphics[width=\columnwidth]{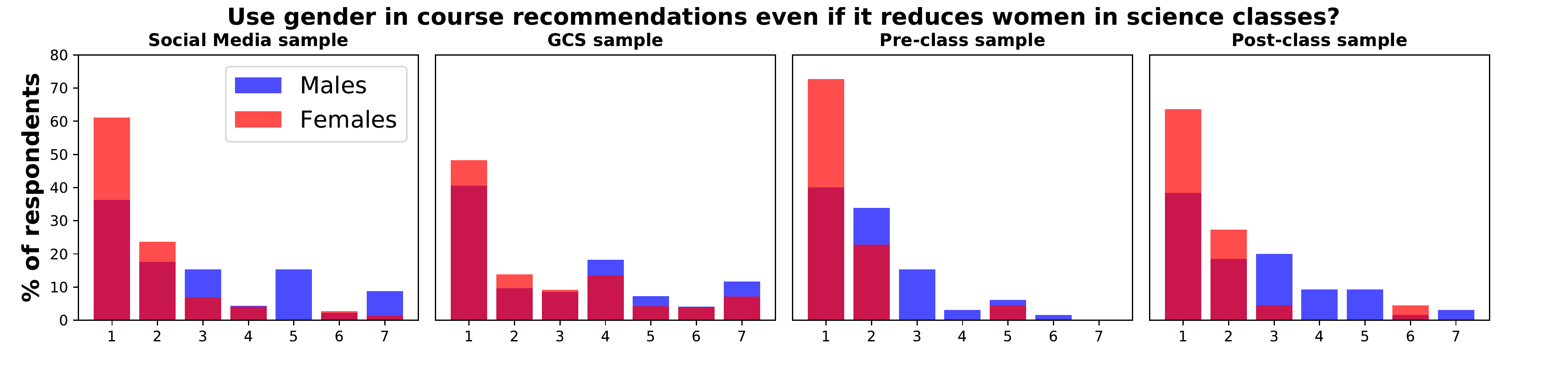}
\caption{Histograms of all survey results for the gender-in-course-recommendations question. Differences between male respondents (blue bars) and female respondents (red bars) are statistically significant in all surveys.} 
\label{fig:combined_gender_results}
\end{figure}

Gender was statistically significantly associated with beliefs about algorithmic fairness ($p < .05$) in all four surveys (Table \ref{tab:combined_gender_results}). Female respondents felt more strongly than male respondents that gender should not be included in an education company's algorithm that recommended classes to students if that would make it less likely that women were recommended science classes, even if it increased the algorithm's accuracy (Figure \ref{fig:combined_gender_results}). The size of this effect varied by population, from 0.5 points in the GCS sample to 1.1 points in the social media sample. (The smaller effect size in the GCS sample may occur in part because respondents were more likely to be in a hurry or to not understand the question, introducing noise). Even in the GCS sample, however, a meaningful gap emerged: male respondents were, for example, 41\% likely to give an answer of 4 or above, as opposed to only 29\% for female respondents, an odds ratio of 1.7. 

Male respondents were also statistically significantly more likely than female respondents to weight maximizing accuracy over minimizing racial disparities in criminal risk prediction in the social media sample and pre-class survey (second row of Table \ref{tab:combined_gender_results}). Results were directionally similar in the post-class survey, but the difference was slightly smaller and not statistically significant ($p = .11$). Male respondents also favored computer algorithms over human judges more strongly than did female respondents in all samples (third row of Table \ref{tab:combined_gender_results}), although this difference was not statistically significant. 

We found no significant associations between a respondent's academic background and their responses to algorithmic fairness questions. (While gender was associated with academic background in the social media sample -- 29\% of computer scientists were female, as opposed to 52\% of humanities respondents and 49\% of other science/math respondents -- it was not in the computer science class sample. We verified that gender effect sizes and $p$-values did not change significantly in any sample when we controlled for academic background). Due to our relatively small sample size in the social media and computer science class samples, lack of statistical significance should not be taken as proof that no effect exists.

\subsection{Can views on algorithmic fairness be changed?}

\begin{figure}[h!]
\centering
\includegraphics[width=\columnwidth]{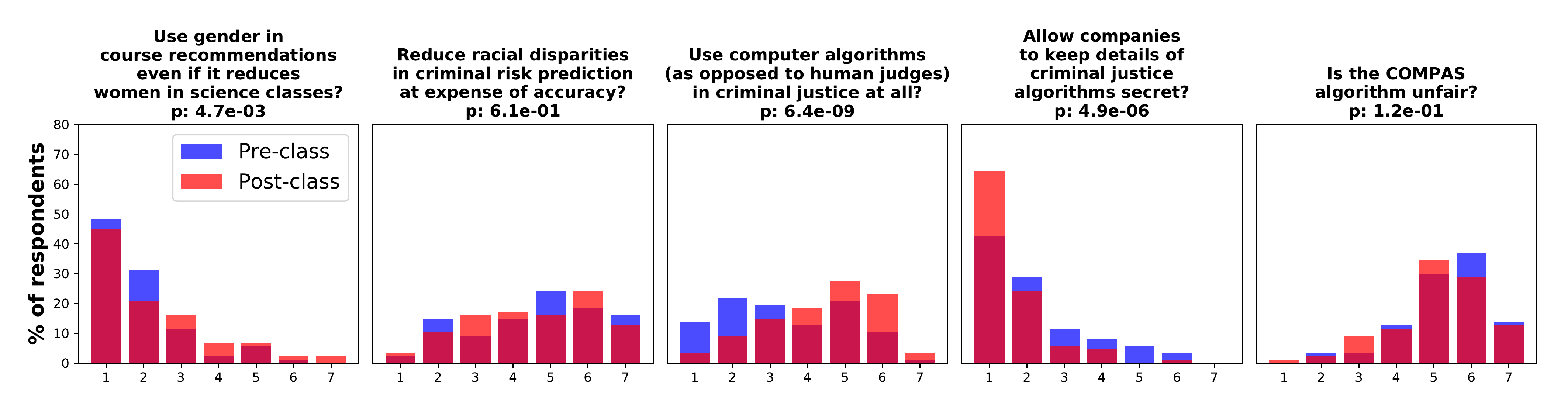}
\caption{Histograms of all pre-class results (blue bars) and post-class results (red bars). $p$ values are calculated using a two-tailed paired t-test.} 
\label{fig:combined_pre_post_results}
\end{figure}

We next assessed, using the pre- and post-class surveys of the students in the undergraduate computer science classes, whether views on algorithmic fairness were changed by lecture and discussion. In Figure \ref{fig:combined_pre_post_results} we show the histograms of pre- and post-class survey results. After lecture and discussion, students were statistically significantly more likely to: 

\begin{enumerate}
\item \textbf{Favor including gender as a feature} in a course-recommendation algorithm, even if it would increase gender disparities. This effect (a surprise to the lecturer) may in part occur because of the discussion in lecture of the weaknesses of ``fairness through blindness'' \cite{dwork2012fairness}: even if a sensitive feature is not included in an algorithm, correlated features may provide information about the sensitive feature. 
\item \textbf{Favor algorithms as opposed to human judges}. Students were swayed by the idea that judges would have the same problems as algorithms but, in addition, could make worse and noisier predictions. 
\item \textbf{Favor algorithmic transparency}. In lecture, algorithmic transparency was presented as a principle academics fairly universally agreed on. Several students commented that, after hearing the lecture, they were more open to using algorithms in criminal justice as long as they were transparent. 

\end{enumerate}

Students also became slightly less convinced that the COMPAS algorithm discussed in the ProPublica article was unfair, but the change in their views was smaller and not statistically significant. This is somewhat surprising because much of the lecture discussed the problems in ProPublica's original analysis---specifically, that there are multiple competing definitions of fairness which cannot be satisfied simultaneously. Students were open to and comprehended this idea---``[The ProPublica article] was very one sided, and this lecture provided a helpful counterpoint" one student commented, echoing several others---but nonetheless did not dramatically revise their views on the fairness of the algorithm. It may be easier to change people's views on objective matters---such as whether algorithms have better predictive performance than human judges---than on more subjective questions of which fairness definitions are best. 

Students were also asked whether any of their views on algorithmic fairness had changed. 74\% of respondents said at least one of their opinions had slightly changed; 17\% said at least one opinion had significantly changed, 1\% said at least one had totally changed, and 8\% said their opinions were exactly the same as before. 

Although discussion can change views on algorithmic fairness, it does not automatically increase consensus. For example, the spread (as measured by both standard deviation and entropy) of the answers to the gender-in-course-recommendations question increased after class, and the spread of the COMPAS fairness question also slightly increased. In both cases, a fairly strong pre-class consensus was less strong after class, possibly reflecting students' greater appreciation for the complexities at play. (``Algorithmic fairness,'' one student commented, ``is much harder than I anticipated''---an admirable summary of the recent academic literature.) Similarly, the gender gaps observed prior to class did not consistently become smaller or disappear after class. 

\section{Discussion}

In this work we make two contributions. First, we show that demographic differences contribute to disagreements about algorithmic fairness. We identify statistically significant and practically meaningful gender discrepancies in beliefs about algorithmic fairness in three separate samples. Our finding is consistent with the results in \cite{grgic2016case}, which identifies evidence of gender differences in beliefs about which features are acceptable to use in criminal risk prediction, although they do not assess statistical significance because their sample is small. \cite{grgic2018human} also identifies substantial disagreement in beliefs about which features are acceptable to use in criminal risk prediction (though they do not assess gender disparities). 

We further show that these differences persist even after an hour-long lecture and discussion, indicating that they are not just the transient products of ignorance or lack of reflection. Future work should also seek to measure \emph{racial} discrepancies in beliefs about algorithmic fairness, a topic of pressing importance given the racial disparities in algorithmic decisions. 

Demographic differences in beliefs about algorithmic fairness have two implications. First, some disagreements about algorithmic fairness may stem from fundamental aspects of background or life experience, and as such have more subjective solutions than purely technical problems. Second, our results support the frequently made argument that the demographics of algorithm designers can affect their algorithmic decisions \cite{clark2016artificial,crawford2016artificial,diversitycrisisai,house2016preparing}. The need for greater diversity among computer scientists is consequently pressing. It is worrisome that the race and gender groups least likely to be involved in algorithmic discussions are also the groups often harmed by algorithmic disparities. If demographics predict how we believe algorithms should behave, we need our algorithm designers to be more demographically representative if algorithms are to serve the will of the whole population. It has long been argued that such ``descriptive representation'' is desirable among democratically elected representatives \cite{mansbridge1999should}; it may be even more important among algorithm designers as a way of protecting disadvantaged groups. Disadvantaged groups can vote against politicians whose policy decisions adversely affect them; disadvantaged groups adversely affected by algorithms often have no such recourse. 

As a second contribution, we show that views on algorithmic fairness can be changed by lecture and discussion. Students in undergraduate computer science classes became more likely to emphasize transparency, more open to using algorithms rather than using judges, and less convinced that gender should not be used as a feature even if it increased disparities. (Of course, the effect of lecture and discussion on student views will vary depending on the focus of the lecture---a topic for future study.) To our knowledge, this is the first quantitative evidence that views of algorithmic fairness can be changed. This finding implies, for example, that even if the public is initially reluctant to use algorithms in criminal justice, that belief is potentially malleable. 

At the same time, discussion should not be taken as a panacea for disagreement. We find that gender gaps persist after discussion, the spread of answers does not consistently decrease, and that most respondents changed their views only slightly. Views on whether the COMPAS algorithm is fair did not shift substantially after lecture, potentially indicating that views on what fairness means are harder to shift than views on more objective questions like whether algorithms have more accurate predictive performance than human judges \cite{kleinberg2017human}. 

Taken together, our two findings suggest that consensus on some algorithmic fairness questions may be difficult to achieve. Differences in demographic backgrounds contribute to disagreement, and hour-long discussions do not necessarily reduce it. While disagreement continues, the demographic differences we observe imply that we should strive for greater diversity among algorithmic decision-makers. If we cannot achieve perfect agreement, we must at least try to ensure that all voices are heard.

\clearpage 
\bibliographystyle{abbrv}
\bibliography{bibliography}

\end{document}